\documentclass[%
 reprint,
 amsmath,amssymb,
 aps,
]{revtex4-1}

\usepackage{graphicx}
\usepackage[space]{grffile}
\usepackage{latexsym}
\usepackage{textcomp}
\usepackage{longtable}
\usepackage{booktabs,array,multirow}
\usepackage{amsfonts,amsmath,amssymb}
\usepackage{natbib}
\usepackage{url}
\usepackage{hyperref}
\hypersetup{colorlinks=false,pdfborder={0 0 0}}
\usepackage{etoolbox}
\newif\iflatexml\latexmlfalse
\AtBeginDocument{\DeclareGraphicsExtensions{.pdf,.PDF,.eps,.EPS,.png,.PNG,.tif,.TIF,.jpg,.JPG,.jpeg,.JPEG}}

\usepackage[utf8]{inputenc}
\usepackage[english]{babel}

\begin{document}

\title{Quantum Beat Photoelectron Imaging Spectroscopy of Xe in the VUV}

\author{Ruaridh Forbes}
\affiliation{Department of Physics, University of Ottawa, 150 Louis Pasteur, Ottawa, ON, K1N 6N5, Canada}
\affiliation{Department of Physics and Astronomy, University College London, Gower Street, London, WC1E 6BT, United Kingdom}

\author{Varun Makhija}
\email{vmakhija@uottawa.ca}
\affiliation{Department of Physics, University of Ottawa, 150 Louis Pasteur, Ottawa, ON, K1N 6N5, Canada}

\author{Jonathan Underwood}
\affiliation{Department of Physics and Astronomy, University College London, Gower Street, London, WC1E 6BT, United Kingdom}

\author{Albert Stolow}
\affiliation{Department of Physics, University of Ottawa, 150 Louis Pasteur, Ottawa, ON, K1N 6N5, Canada}
\affiliation{Department of Chemistry, University of Ottawa, 10 Marie Curie, Ottawa, Ontario, K1N 6N5, Canada}
\affiliation{National Research Council of Canada, 100 Sussex Drive, Ottawa, Ontario K1A 0R6, Canada}

\author{Iain Wilkinson}
\affiliation{National Research Council of Canada, 100 Sussex Drive, Ottawa, Ontario K1A 0R6, Canada}
\affiliation{Methoden der Materialentwicklung, Helmholtz-Zentrum Berlin für Materialien und Energie GmbH, Hahn-Meitner-Platz 1, 14109 Berlin, Germany}

\author{Paul Hockett}
\email{paul.hockett@nrc.ca}
\affiliation{National Research Council of Canada, 100 Sussex Drive, Ottawa, Ontario K1A 0R6, Canada}

\author{Rune Lausten}
\affiliation{National Research Council of Canada, 100 Sussex Drive, Ottawa, Ontario K1A 0R6, Canada}



\selectlanguage{english}
\begin{abstract}
Time-resolved pump-probe measurements of Xe, pumped at 133~nm and probed at 266~nm, are presented. The pump pulse prepared a long-lived hyperfine wavepacket,  in the Xe $5p^5(^2P^{\circ}_{1/2})6s~^2[1/2]^{\circ}_1$ manifold ($E=$77185~cm$^{-1}=$9.57~eV). The wavepacket was monitored via single-photon ionization, and photoelectron images measured. The images provide angle- and time-resolved data which, when obtained over a large time-window (900~ps), 
constitute a precision quantum beat spectroscopy measurement of the hyperfine state splittings. Additionally, analysis of the full photoelectron image stack provides a quantum beat imaging modality, in which the Fourier components of the photoelectron images correlated with specific beat components can be obtained. This may also permit the extraction of isotope-resolved photoelectron images in the frequency domain, in cases where nuclear spins (hence beat components) can be uniquely assigned to specific isotopes (as herein), and also provides phase information. The information content of both raw, and inverted, image stacks is investigated, suggesting the utility of the Fourier analysis methodology in cases where images cannot be inverted. 
\end{abstract}%

\maketitle

\textit{Publication history}
\begin{itemize}
\item \href{https://www.authorea.com/users/71114/articles/188337-quantum-beat-photoelectron-imaging-spectroscopy-of-xe-in-the-vuv}{Original manuscript (Authorea)}, Feb. 2018.
\item arXiv, March 2018.
\end{itemize}

\textit{Supplementary material} 

OSF project \href{https://osf.io/ds8mk/}{\textbf{Quantum Beat Photoelectron Imaging Spectroscopy of Xe in the VUV}}, DOI: \href{http://dx.doi.org/10.17605/OSF.IO/DS8MK}{10.17605/OSF.IO/DS8MK}


\section{Introduction}
Quantum beat spectroscopy (QBS) provides a time-domain route to high-resolution spectroscopic measurements \cite{Haroche_1973}. In a standard scheme a narrow wavepacket  (few-state superposition) is prepared, time-domain measurements are obtained, and Fourier analysis of the signal provides the high-resolution frequency-domain information sought. 
More generally, quantum beat spectroscopy can be regarded as a subset of generalized wavepacket methods \cite{Tannor2007}, with the specific requirement that sufficient wavepacket revivals are present in the observed temporal window to provide frequency-domain information. The applicability, and details, for a given case will therefore depend on experimental factors - e.g. time-resolution, wavepacket preparation - and intrinsic systems properties - e.g. density of states, lifetimes \cite{Hack_1991, Blum_1996}. Wavepackets comprised of fine and hyperfine levels in rare gases and alkali atoms are a notable application of QBS, since the lifetimes and level spacings are concomitant with ns and ps experimental time-scales \cite{Haroche_1973, Arimondo_1977}. 



In order to obtain time-domain data with good signal to noise, an observable which responds to the wavepacket dynamics is required. In many cases, photoelectron angular distributions (PADs) provide a sufficient observable, which may be much more  sensitive to underlying dynamics than photoelectron yields or energy spectra alone. 
In particular, the PADs are sensitive to the angular momentum couplings in the system; for the case of hyperfine interactions, this sensitivity has been investigated extensively by Berry and co-workers for sodium and lithium, prepared with ns pulses \cite{Strand_1978,Hansen_1980,Chien_1983}. In this pioneering atomic ionization work, PADs were measured (in a plane perpendicular to the light propagation) for different time-delays and pump-probe polarization geometries, and the data obtained was sufficient to allow extraction of the photoionization matrix elements (partial-wave magnitudes and phase shifts), and an effective nuclear spin depolarization parameter, related to the hyperfine coupling. The analysis of the temporal evolution was hindered by both the experimental difficulty of recording PADs, rendering it feasible to record PADs at only a few pump-probe delays, and the fact that the pulse durations were non-negligible compared to the hyperfine precession (i.e. the characteristic time-scale of the coherently prepared hyperfine  wavepacket) \cite{Hansen_1980}. Nonetheless, the requisite theory and a detailed physical understanding of the hyperfine interaction was developed (see also refs. \cite{Greene1982,Klar1982} for related theory work). Subsequent work by various investigators explored this topic further, for example Bajic $\textit{et al.}$ \cite{Bajic_1991}, who investigated angle-resolved multi-photon ionization in Kr and Xe, and  Reid $\textit{et al.}$ \cite{Reid_1994}, who explored hyperfine depolarization in NO via time and angle-resolved photoelectron measurements; in both cases the work again highlighted the sensitivity of PADs to hyperfine interactions.


Xenon has two naturally occurring isotopes with non-zero nuclear spin ($^{129}$Xe ($I=1/2$), $^{131}$Xe ($I=3/2$)), and the hyperfine level structure has been well-studied in the energy domain at a range of energies with a variety of methods, including fluorescence \cite{Meier_1977}, saturated-amplification \cite{Cahuzac_1975} and photoionization \cite{Brandi_2001,Paul_2005,W_rner_2005}; for a more comprehensive overview see ref. \cite{D_Amico_1999}. A range of high (energy) resolution ionization experiments incorporating the preparation of high-n Rydberg states have been performed to probe hyperfine splittings in Ryberg manifolds, including autoionizing regions, and in the cation \cite{Sch_fer_2010,Sukhorukov_2012}. Recently, photoion-photoelectron coincidence experiments provided isotopically-resolved (hence $I$-resolved) PADs from the $(^2P^{\circ}_{3/2})5d~^2[3/2]^{\circ}_1$ state, following VUV excitation at 10.4~eV (83876~cm$^{-1}$) 
\cite{O_Keeffe_2013}; subsequent work included detailed theoretical analysis, and the first determination of the hyperfine couplings in this wavelength range \cite{Gryzlova_2015}. 

In this work, broadband femtosecond VUV pulses ($\lambda \approx$133~nm, $E_{h\nu}\approx$9.32~eV~$=$75188~cm~$^{-1}$, $\Delta\lambda\approx$1.7~nm, $\tau\approx$80~fs) were used to coherently prepare hyperfine states in the Xe$(^2P^{\circ}_{1/2})6s~^2[1/2]^{\circ}_1$ manifold\footnote{In Racah notation, where the core is given by the term symbol ($^{2S+1}L_J$) and the excited electron is defined by $nl~^{2S+1}[K]_{J_e}$, where $K~=~J+l$ and $J_e~=~K+s$.} ($E=$77185~cm~$^{-1}=$9.57~eV,  \href{https://physics.nist.gov/PhysRefData/Handbook/Tables/xenontable5.htm}{NIST value} \cite{Kramida2018}, adapted from ref. \cite{Brandi_2001}), which were subsequently ionized with UV pulses ($\lambda=$266.45~nm, $E_{h\nu}=$4.653~eV~$=$37530~cm~$^{-1}$, $\Delta\lambda=$3~nm, $\tau=$50~fs). The experimental set-up, and details of the VUV generation are presented in Sect. \ref{sect:expt}. In the experiments, photoelectron images were obtained as a function of VUV-UV delay, over a 900~ps temporal window (Sect. \ref{sect:Images}). As discussed above, temporal modulations in the PADs provide an observable sensitive to the underlying wavepacket dynamics, and this data constitutes a QBS measurement. Determination of the hyperfine splittings, and coupling constants, from the photoelectron data is discussed in Sect. \ref{sect:QBS}. The data additionally suggests a \textit{quantum beat imaging} methodology, in which beat-frequency resolved photoelectron images, and associated phase information, may be obtained: this is discussed in Sect. \ref{sect:QBI}.

The raw experimental data, data processing scripts, and additional analysis notes to accompany this manuscript can \href{https://osf.io/ds8mk/}{be found online via an OSF repository}, 
see Sect. \ref{sec:SuppMat} for further details.

\section{Experiment\label{sect:expt}}

\subsection{VUV-UV photoelectron imaging}
The experimental set-up used for the VUV-UV pump-probe photoelectron velocity-map imaging (VMI) measurements at 133~nm reported herein was almost identical to that previously reported for work using 160~nm radiation \cite{Forbes_2017}, and the reader is referred to that work for further details of the experimental apparatus beyond the outline sketched here. A similar VUV-UV pump-probe VMI experimental configuration has also been previously reported by Suzuki (see ref. \cite{Suzuki_2014} for a summary); for further general discussion of VUV laser spectroscopies in both time and frequency domain precision spectroscopy experiments see, e.g., ref. \cite{Eikema2011}.

\begin{figure*} 
\begin{center}
\includegraphics[width=1\textwidth]{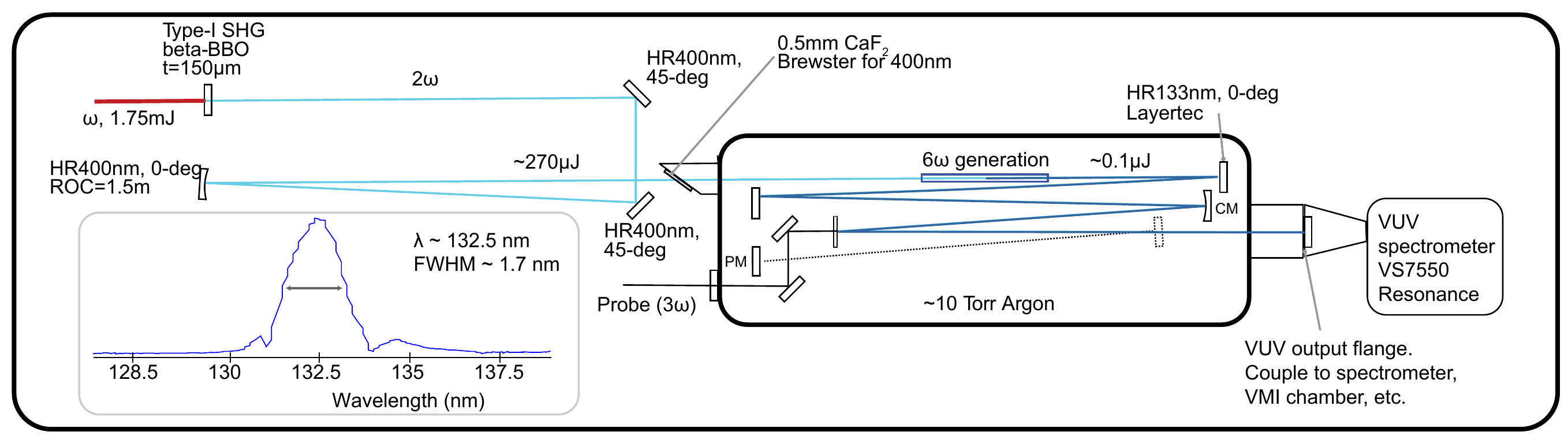}
\caption{{~ Layout of the 6\(\omega\) generation chamber, configured to
record the VUV spectrum. For photoelectron experiments, the VUV
spectrometer was replaced with a VMI spectrometer. HR: high-reflector,
CM: curved mirror. Inset shows a typical~ 6\(\omega\)~spectrum,
recorded with a VUV spectrometer (Resonance VS7550, uncalibrated, see
main text for details). ~
{\label{574557}}%
}}
\end{center}
\end{figure*}

Briefly, the optical chain was initiated by a standard amplified titanium-sapphire laser system (Coherent Legend-Elite Duo), which provided 35~fs pulses at
795~nm at 1~kHz. A 2.5~mJ component of the total laser output, 7.5~mJ, was utilized in the experiments presented here. The beam was placed through a 70:30 beamsplitter to provide the pump and probe arms, respectively. The 1.75~mJ is frequency doubled using a 150~$\mu$m Beta Barium Borate ($\beta$-BaB$_{2}$O$_{4}$, $\beta$-BBO), to provide 397.5~nm (2$\omega$) pulses with an estimated pulse duration of 40~fs. The 2$\omega$ light was then separated from the fundamental by using dielectric mirrors. The 2$\omega$ pulses were the focused (ROC = 1.5~m) into an argon-filled gas cell, 
where the 2$\omega$ pulse was frequency tripled by a six-wave mixing process, as described by Trabs and coworkers \cite{peter2015, Trabs_2016}. The 6$\omega$ femtosecond pulse was then separated from the driving field, and refocused, using 0 deg. reflections from dielectric mirrors centred at 133nm (Layertec GmbH). The optical layout of the VUV generation chamber, and a representative spectrum of the generated radiation, is shown in figure \ref{574557}. The spectrum was obtained with a VUV spectrometer (\href{http://resonance.on.ca/vs7550-vuv-uv-mini-spectrometer/}{Resonance VS7550}) in the configuration as shown in fig. \ref{574557}. The recorded spectrum is uncalibrated, but expected to be centred at $\lambda~=~$132.5~nm based on the driving field wavelength, and stable as a function of driving field intensity \cite{peter2015, Trabs_2016}. Although this is somewhat to the red of the target pump transition to populate states in the $(^2P^{\circ}_{1/2})6s~^2[1/2]^{\circ}_1$ manifold lying at $E~=~$77185~cm$^{-1}$ ($\lambda~=~$129.6~nm), significant resonance-enhanced photoelectron signal was observed in the VMI measurements detailed below. This indicates that the wings of the pulse contained sufficient flux to drive the pump transition, and/or a slight blue-shift of the spectrum from the expected central wavelength.

For photoelectron imaging experiments, the VUV spectrometer was replaced with a VMI spectrometer \cite{Eppink_1997}, separated from the VUV generation chamber by a minimal thickness (0.5~mm) CaF$_2$ window (Crystran). For pump-probe measurements, the remaining 0.75mJ of the fundamental transmitted through the beamsplitter was delayed using a motorized stage (Newport XML210), frequency tripled in two $\beta$-BBO crystals (see ref. \cite{Forbes_2017} for further details), and then recombined with the 6$\omega$ pulse, in a collinear geometry.

The pump and probe beams were focused through a set of baffles, to minimize signals from scattered VUV and UV light, into the interaction region of a VMI spectrometer. An atomic beam of Xe was generated in a separate source chamber by expanding a 7.5\% mix of Xe (Praxair Canada Inc., 5N purity) seeded in He (BOC GAZ, 5N purity) through an \href{https://sites.google.com/site/evenlavievalve/}{Even-Lavie valve} \cite{Even_2000}, operating at 1~kHZ and held at 40~$^{o}$C, at a pressure of 20~psi (138~kPa). The supersonic expansion was skimmed, yielding an atomic beam with an estimated diameter of $\sim$1~mm, before entering the interaction along the VMI Time-of-Flight (ToF) axis. The VMI spectrometer consisted of a three-stage, open aperture repeller electrode system, and a ten-element Einzel lens stack \cite{Forbes_2017}. 
Accelerated electrons were detected using an MCP-Phosphor detector setup and photoelectron images recorded by relay-imaging the phosphorescence at a CCD camera (Thorlabs DC210). A cross-correlation of $\tau_{xc}=$170~fs (FWHM) in the interaction region of the VMI spectrometer was determined by resonantly enhanced two-color ionization of xenon. The extended duration of the cross-correlation (as compared to the optimal transform-limited cross-correlation value of ca. 60~fs for these pulses) is attributed primarily to dispersion of the 6$\omega$ pulse in the CaF$_2$ window. In future work, a switch to an LiF window, combined with upstream dispersion compensation, should enable transform-limited 6$\omega$ pulses (ca. 40~fs) in the interaction region of the spectrometer.

Time-resolved photoelectron images were recorded for pump-probe delays between $t=$\selectlanguage{english}-70~ps and $t=$+890~ps in steps of 10~ps. Here, negative delay refers to the situation where the 3$\omega$ pulse arrives before the 6$\omega$. The time-resolved photoelectron signals were constructed from the measured photoelectron images in the following manner. At each time-delay of the pump and probe, two photoelectron images where recorded: one with the gas pulse temporally overlapped with the two lase pulses; and one without the gas pulse (to account for ionization due to background contaminant gases and scattered light signals associated with the 6$\omega$ pulse). The ``no gas" signal was then subtracted from the gas pulse data to obtain background gas and scatter free images. It is of note that no one-color photoelectron counts, bar scatter light signals, were observed with the 6$\omega$ pulse. Additionally, negligible counts with 3$\omega$ pulse were observed with the pulse energies and MCP/phosphor voltage settings employed during the collection of the images. In total, 25 scans of the pump-probe delay window were performed. Energy to pixel calibration was achieved by recording 3-photon ionization of Xe at 266~nm under the same VMI focusing conditions utilized in the experiment.


\subsection{\label{sec:dataAnalysis}Data processing}
The energy and time-dependent photoelectron distributions can be defined in the usual way, in terms of a spherical harmonic expansion (see, e.g., ref. \cite{Reid_2003}) 
\begin{equation}
I(\theta,\phi,E,t)=\sum_{L,M}\beta_{L,M}(E,t)Y_{L,M}(\theta,\phi)
\end{equation}
where $E$ is the photoelectron kinetic energy and $t$ the pump-probe delay; $\beta_{L,M}$ are the anisotropy parameters, which can be related to the photoionization dynamics of the system. For cylindrically-symmetric distributions, only $M=0$ terms are non-zero, and the $\phi$ coordinate is redundant: this is the case for the measurements reported herein, and these terms are subsequently omitted in this manuscript. For 2D imaging of cylindrically-symmetric 3D distributions, in which the 3D photoelectron distribution is projected onto the detector in the measurements (sometimes termed ``crush" imaging), the symmetry of the projection enables 2D slices from the original 3D photoelectron distributions to be reconstructed using standard inversion techniques \cite{ImagingBook_2003}. 
To reconstruct the slices, and determine the $\beta_{L}(E,t)$ metrics, the photoelectron images were processed using \href{https://github.com/e-champenois/CPBASEX}{cpBasex}, which implements the pBasex inversion method \cite{Garcia2004}. This method uses a fitting methodology, and provides both inverted (or ``slice") images, hereafter denoted by the coordinate system $(x_i,y_i,t)$, and the associated $\beta_{L}(E,t)$ expansion parameters directly. For the distributions considered herein, from a two-photon process with cylindrical symmetry, the only non-zero parameters are $L~=~0,2,4$.  (For non-cylindrically symmetric distributions, e.g. distributions arising from non-parallel pump-probe polarization geometries, direct inversion is not possible, although other methods - for instance tomographic reconstruction - can be applied with some additional experimental effort. For further general discussion on charged particle imaging and reconstruction, see ref. \cite{ImagingBook_2003}, for recent discussion in the context of 2D and 3D metrology techniques, including photoelectron tomography, see, for instance, ref. \cite{Hockett_2015} and references therein.)

Images were defined and processed via a variety of analysis protocols, including per-scan and scan-summed, and with and without image symmetrization. The processing of images from each experimental scan provided a way to estimate statistical uncertainties, while scan-summed and symmetrized images provided the best signal-to-noise, hence highest resolution, dataset. Other selections, e.g. summation over a sub-set of the scans, choice of a single quadrant from the photoelectron images, and so forth, provided additional validation and cross-checks on the extracted data. In the results reported herein, the scan-summed dataset was the main focus of the analysis, and statistical (1$\sigma$) uncertainties were determined from analysis on a per-scan basis  (see also Sect. \ref{sec:SuppMat}).




\section{Results and Discussion}

\subsection{Photoelectron images\label{sect:Images}}
\begin{figure} 
\begin{center}
\includegraphics[width=1.00\columnwidth]{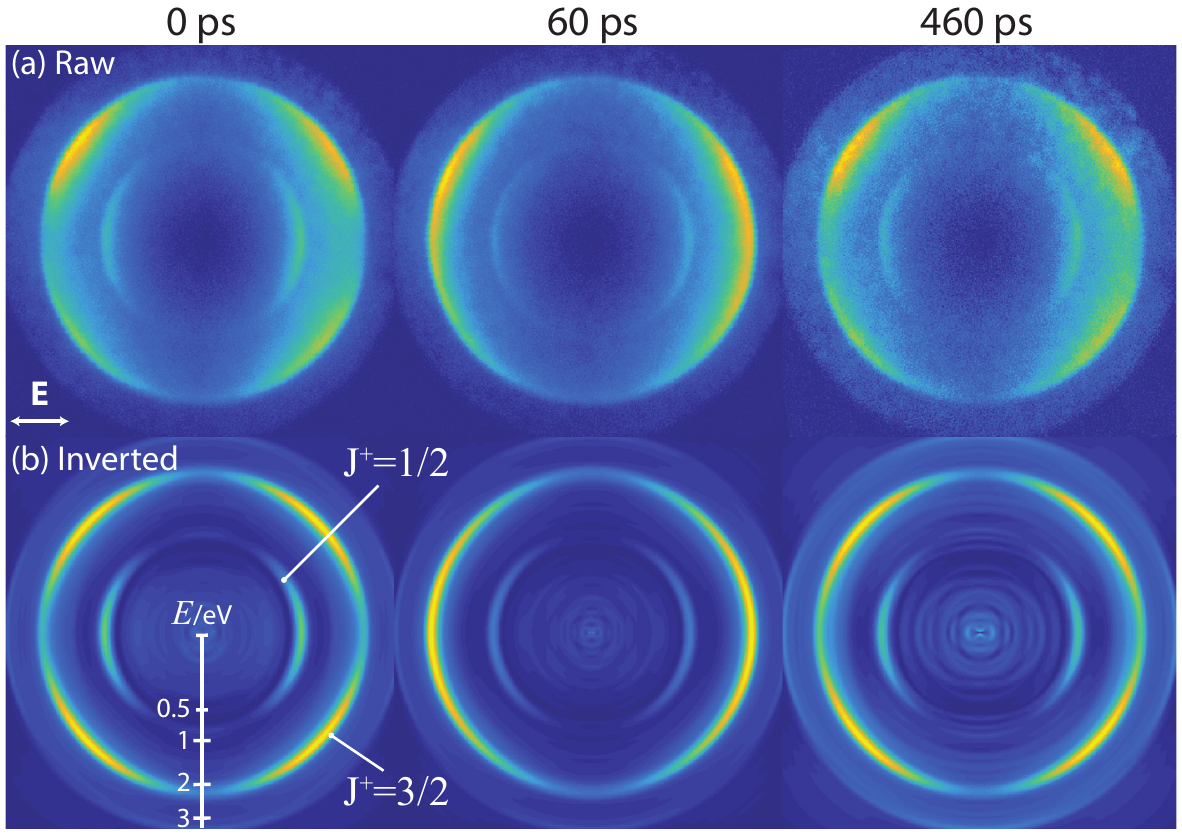}
\caption{{Example photoelectron images. (a) Raw images, summed over experimental
cycles. The electric field polarization vector was horizontal, as
indicated. (b) The corresponding inverted images. Photoelectron bands
are labelled by the final cation states,~~\(J^+=\frac{1}{2}\)
and~\(J^+=\frac{3}{2}\), corresponding to electron
energies~\(E=\) 0.8 eV and~\(E=\) 2.1 eV
respectively. ~Scale bar shows pixel (\(\ \propto\) velocity)
to~\(E\) conversion. Colour maps were normalised to the
maximum signal intensity for each image (arb. units). The differences
between quadrants in the raw images, which should be identical by
symmetry, is ascribed primarily to detector inhomogeneities, although
other experimental factors may also contribute.
{\label{772944}}%
}}
\end{center}
\end{figure}

Figure \ref{772944}(a) illustrates typical raw (un-symmetrized) photoelectron imaging results at various pump-probe time delays. The images show a dramatic change in the angular dependence of the outer ring as a function of pump-probe delay, $t$, with clear switching from a four-lobed to two-lobed structure. At a total photon energy of 14.24~eV (114850~cm~$^{-1}$), there is 2.1~eV of excess energy above the first ionization threshold at 12.13~eV (97833~cm$^{-1}$, values from  \href{https://physics.nist.gov/PhysRefData/Handbook/Tables/xenontable1.htm}{NIST}, ref. \cite{Kramida2018}, adapted from refs.  \cite{Knight_1985,Brandi2001}), and two final $J^+$ states can be populated in the ion: 
\begin{itemize}
\item Xe$^+(5p^5)^2P^{\circ}_{3/2}$ (ground state)
\item Xe$^+(5p^5)^2P^{\circ}_{1/2}$ (1.3~eV, 10537~cm~$^{-1}$)
\end{itemize}
The outer photoelectron band (maximum photoelectron energy) in the images is therefore correlated with formation of the cation with $J^+=3/2$, while the inner ring correlates with the $J^+=1/2$ spin-orbit excited state; the corresponding photoelectron energies (band centres) are $E$=2.1~eV and $E$=0.8~eV, respectively. Inverted (slice) images are also shown in Figure \ref{772944}(b) for reference. 

\begin{figure*} 
\begin{center}
\includegraphics{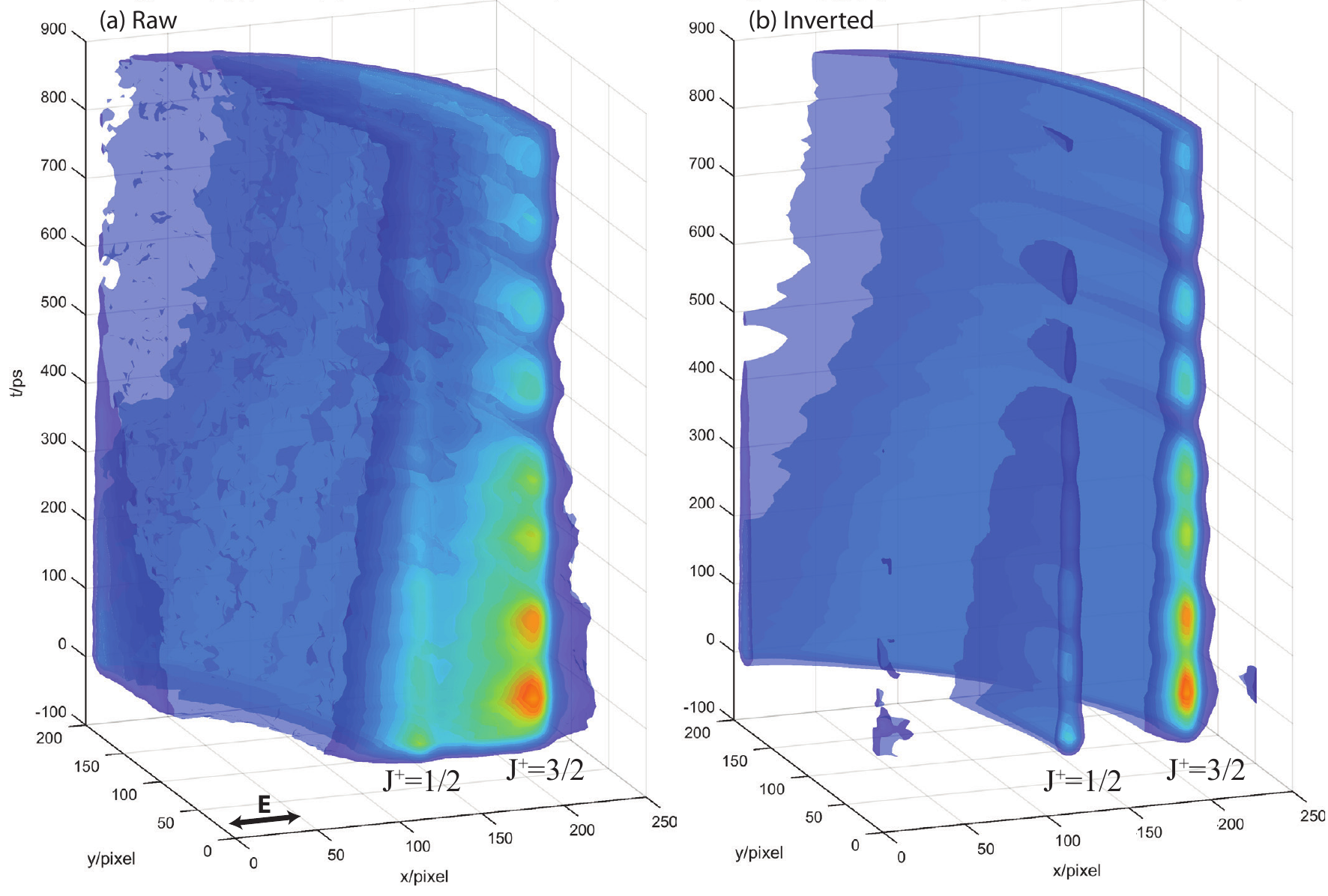}
\caption{{Full image volumes, shown for a single quadrant. (a) Raw data
\(\left(x,y,t\right)\). (b) Inverted images \(\left(x_i,y_i,t\right)\). Isosurfaces
show 10 - 90\% photoelectron signal. The full data volume was
down-sampled and smoothed for these renderings.
{\label{405092}}%
}}
\end{center}
\end{figure*}

A visualisation of the full $(x,y,t)$ volume is shown in figure \ref{405092}(a), and the inverted image volume $(x_i,y_i,t)$ in figure \ref{405092}(b). In these renderings only the top right quadrant of the images is included, and 10 isosurfaces are shown, spaced over 10~-~90\% photoelectron yield (normalised to the maximum volume element). The renderings give a sense of the full dataset: temporal oscillations are clearly observed, particularly in the outer photoelectron band, and some aspects of the changing angular distributions can be discerned (this is less apparent in the static renderings, but can be seen more clearly in \href{http://https://osf.io/ds8mk/wiki/home/}{the interactive versions available online}), although are not pronounced.

\begin{figure} 
\begin{center}
\includegraphics[width=1.00\columnwidth]{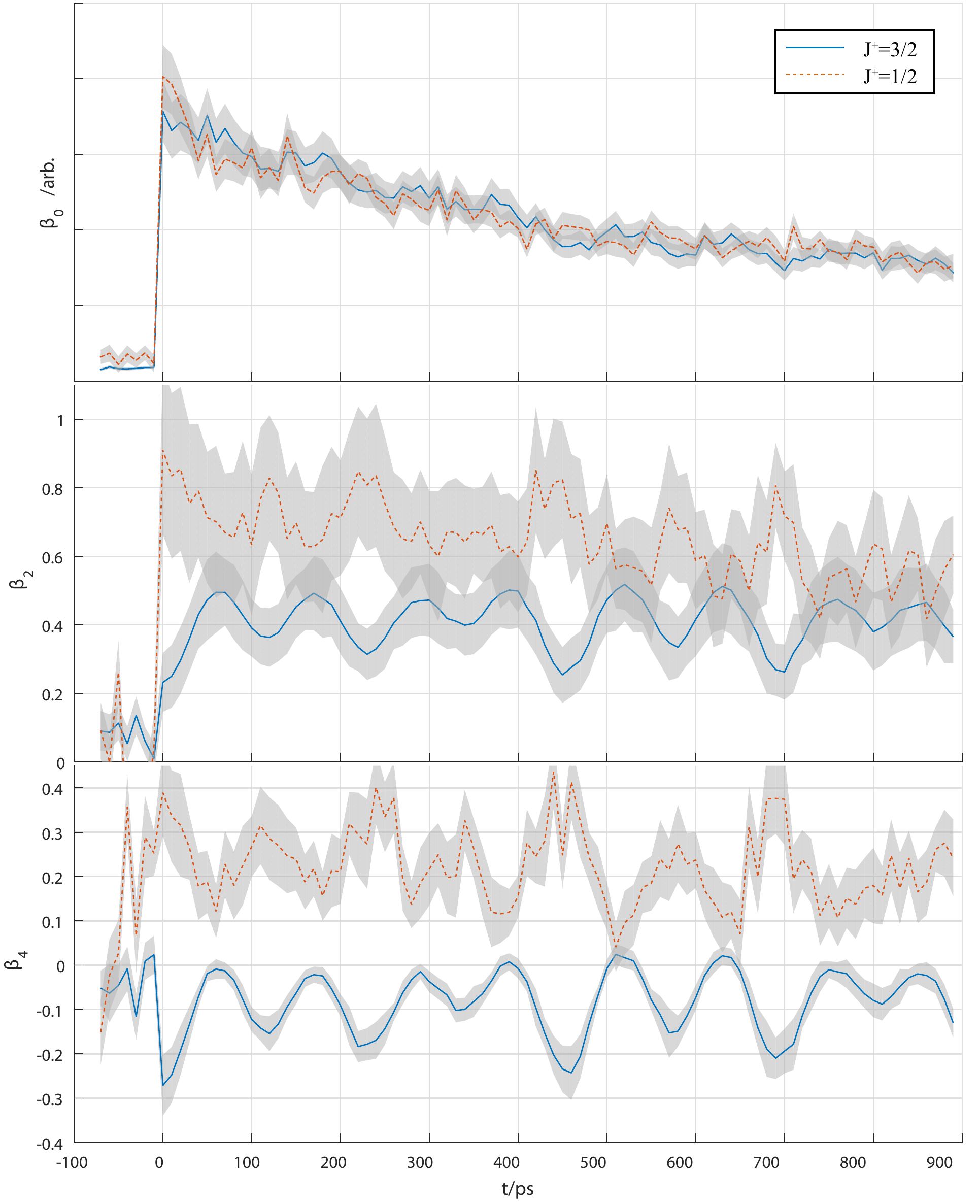}
\caption{{Time-dependent \(\beta_{L}(t)\) parameters, \(J^+=\ \) 3/2
(blue, solid), \(J^+=\) 1/2 (orange, dashed). Uncertainties
show statistical errors (1\(\sigma\)), determined from analysis
of each experimental cycle - see Sect.
{\ref{sec:dataAnalysis}} for details.
{\label{622685}}%
}}
\end{center}
\end{figure}

 A more quantitative picture is obtained by analysis of the results in terms of the characteristic $\beta_{L}(\Delta E,t)$ parameters, as discussed in Sect. \ref{sec:dataAnalysis}. Metrics extracted from the processed images over a small energy range $\Delta E$, defined by the FWHM of the photoelectron features, are illustrated in figure \ref{622685}.  
 It is immediately apparent from the data that the PADs change significantly as a function of pump-probe delay, with clear oscillations apparent in the temporal profiles of $\beta_{2}(t)$ and $\beta_{4}(t)$ for both photoelectron bands. The oscillations are not, however, clearly observed in the photoelectron yield (denoted $\beta_{0}(t)$), which shows a gradual decay with only a hint of the oscillations observed in the $L>0$ terms. This is due to the strong dependence of the PADs on the evolution of the hyperfine wavepacket as a function of time, while the total yields are much less sensitive \cite{Chien_1983, Reid1994}. 
 
 For the inner band ($J^+=1/2$), the lower photoelectron yield results in higher noise in the extracted parameters, but clear quantum beats are still observed. These beats are out-of-phase with the $J^+=3/2$ traces. Empirically, this indicates a sign change in the excited state polarization (alignment tensor) sensitivity of the two ionizing transitions. The sign change is consistent with the treatment of Greene and Zare \cite{Greene1982}, in which a universal alignment function is derived; the sign of this function depends on angular momentum transfer, and changes between $\Delta J=0$ and $\Delta J=\pm1$ ionizing transitions. This behaviour can be considered as analogous to the polarization sensitivity in fluorescence measurements \cite{Haroche1973,Fano1973} although, for the photoionization case, additional terms - including the photoelectron angular momentum - play a role in determining the modulation depth (sensitivity), and the structure of the PADs \cite{Chien_1983,Klar_1982}. 
 


\subsection{Quantum beat spectroscopy: Hyperfine structure\label{sect:QBS}}
\begin{figure} 
\begin{center}
\includegraphics[width=1.00\columnwidth]{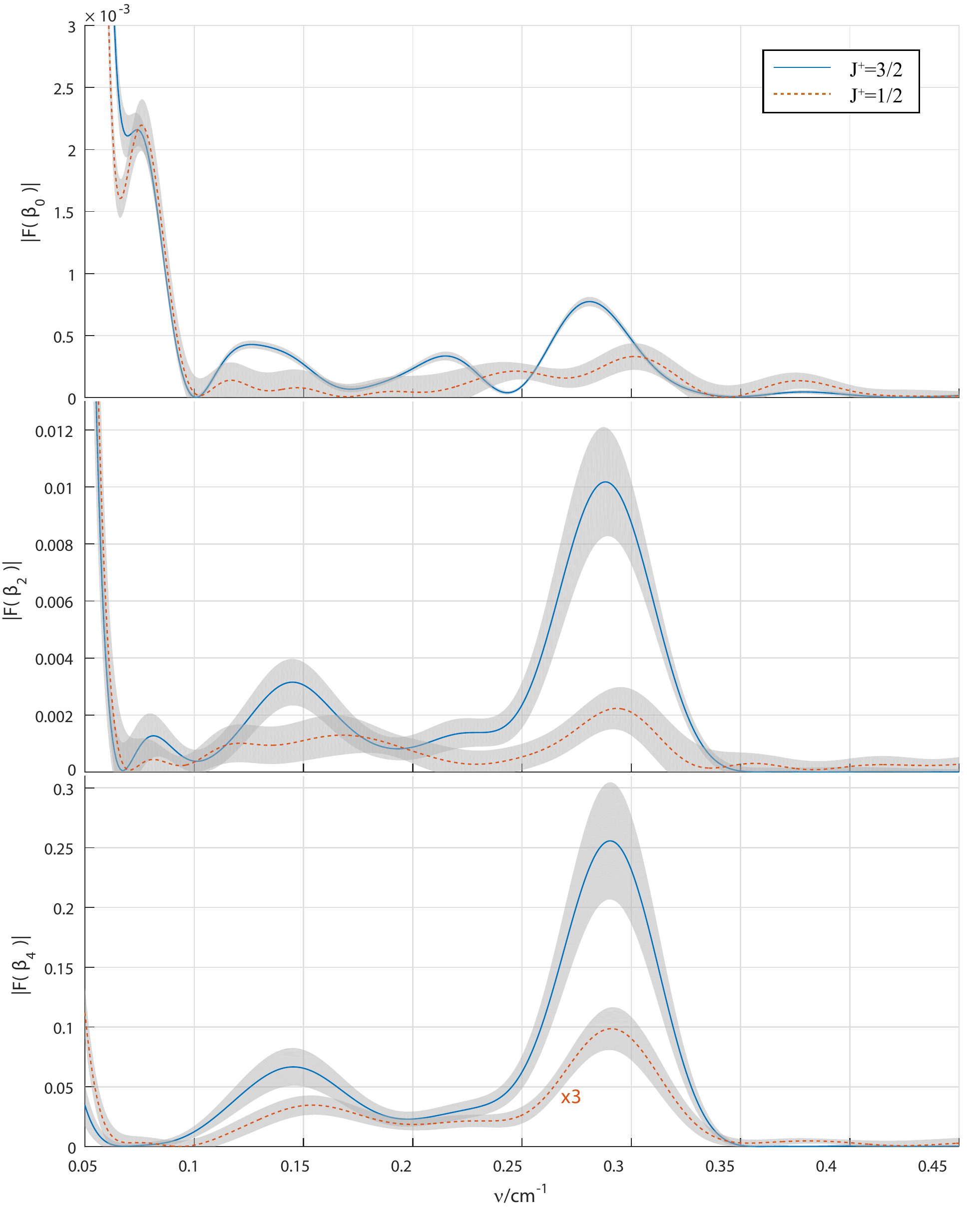}
\caption{{Frequency domain results~\(\beta_L\left(\nu\right)\), obtained via Fourier
transform of the ~\(\beta_{L}(t)\) data shown in
fig.~{\ref{622685}}. Plots show the Fourier power
spectrum for each trace, with statistical uncertainties
(\(1\sigma\)), determined from analysis of each experimental
cycle - see sect.~{\ref{sec:dataAnalysis}} for details.
~
{\label{586171}}%
}}
\end{center}
\end{figure}

Figure \ref{586171} shows Fourier power spectra resulting from the Fourier transform (FT) of the $\beta_{L}(t)$ (Figure \ref{622685}). The corresponding feature positions and uncertainties are listed in table \ref{tab:HFS}. The FTs of both $\beta_{2}(t)$ and $\beta_{4}(t)$ provide the frequency domain parameters $\beta_{2}(\nu)$ and $\beta_{4}(\nu)$, associated with both spin-orbit states of the cation. The major features are located around 0.14~cm~$^{-1}$ and 0.29~cm~$^{-1}$, and minor features around 0.09~cm~$^{-1}$ and 0.22~cm~$^{-1}$ are observed in some channels. The FT of the photoelectron yields ($\beta_{0}(\nu)$) reveal very little frequency structure, apart from the lowest frequency feature.

The frequencies listed in table \ref{tab:HFS} correspond to the peak of the features observed in the FTs from the major ($J^+=3/2$) feature\footnote{Slight shifts in the features for the minor channel ($J^+=1/2$) were observed in some cases, and the statistical uncertainties were significantly larger. Both effects were attributed to the worse signal-to-noise for the minor feature (see figure \ref{622685}).}, and are reported with uncertainties determined from propagation of the 1$\sigma$ errors, extracted by analysis of the each scan, as detailed in sect. \ref{sec:dataAnalysis}. The statistical uncertainties determined in this manner, on the order of $10^{-3}$~cm$^{-1}$, are an order of magnitude better than the absolute frequency limits imposed by the 890~ps temporal window, which corresponds to a lower limit on the observable frequency of  $\nu_{min}~=~3.75\times10^{-2}$~cm$^{-1}$, and defines the resolution of the FT; this limit is also reflected in the feature widths, which are on the same order as $\nu_{min}$. The statistical uncertainties are, however, significantly worse than the absolute experimental frequency accuracy, which is defined by the timing uncertainty of the measurements: in this case, the pump-probe cross-correlation of the laser pulses, $\tau_{xc}~\approx~170$~fs (dispersion limited in the current experiments), which defines a frequency accuracy  $\nu_{xc}~=~6.2\times10^{-6}$~cm$^{-1}$. 
The upper limit on the observable frequency is defined by the temporal sampling step size, $\tau_s~=~10$~ps, which results in $\nu_{max}~=~3.335$~cm$^{-1}$.

From the measurements, the hyperfine coupling constants can be determined by fitting to the usual form (see, e.g., ref. \cite{D_Amico_1999}):
\begin{equation}
\Delta E_{(F,F-1)}=AF+\frac{3}{2}BF\left(\frac{F^{2}+\frac{1}{2}-J(J+1)-I(I+1)}{IJ(2J-1)(2I-1)}\right)
\end{equation}

Where $A$ is the magnetic dipole constant, and $B$ the electric quadrupole constant. Hyperfine constants determined in this manner are reported in table \ref{tab:HFS}, and compared with previously reported values. In one case, $^{131}A$, the constant is comparable to those previously determined within the experimental uncertainty; in the remaining cases, the constants are comparable to those previously determined, but not within the experimental uncertaintites. This may indicate systematic errors in the experiment and/or data analysis; the presence of electric fields (on the order of 10~-~50~Vcm$^{-1}$ in the present case) in the VMI spectrometer is potentially one source of small shifts (sub-cm$^{-1}$) on the measured splittings, particularly since the Stark shifts are $I$ and $F$ dependent \cite{Angel_1968}, and a relatively high-lying manifold is accessed. Detailed exploration of this effect remains for future work, although rough estimates on the scale of the effect for Rydberg states can be made from a hydrogenic model \cite{Zimmerman_1979,Chupka_1993}. In this model, Stark splitting between adjacent levels is given by $\Delta E_s=1.28\times10^{-4}n\varepsilon$, where $\varepsilon$ is the field strength in Vcm$^{-1}$. This indicates $\Delta E_s$=0.04~cm$^{-1}$ for $n=6$ and $\varepsilon$=50~Vcm$^{-1}$, which is significant on the scale of the measured splittings. 
However, it is of note that previous high-resolution studies of higher $n$ manifolds ($n>10$), at higher field strengths of $\varepsilon\sim$100~Vcm$^{-1}$, have neglected such effects \cite{Kono_2013}. The polarizabilities of the relevant states in Xe required to calculate $\Delta E_s$ accurately \cite{Angel_1968} are, to the best of our knowledge, not known.



\begin{table*}
\begin{centering}
\begin{tabular}{|c|c|c|c|c||c|c|}
\hline 
Isotope & $F$, $F'$ & Splitting/cm$^{-1}$ & \multicolumn{2}{c||}{Hyperfine consts.} & \multicolumn{2}{c|}{Literature}\tabularnewline
\hline 
 &  &  & A/MHz & B/MHz & A/MHz & B/MHz\tabularnewline
\hline 
\hline 
129 ($I=1/2$) & 1/2, 3/2 & 0.2863 (5) & -5723 (9) & - & -5808 (2) {[}a{]}, -5806 (4) {[}b{]}, -5799 (9) {[}c{]} & - \tabularnewline
\hline 
131 ($I=3/2$) & 3/2, 1/2 & 0.0855 (10) & 1697 (30) & -8 (7) & 1709.3 (7) {[}a{]}, 1710 (6) {[}b{]}, 1716 (3) {[}c{]} & 30.3 (8) {[}a{]}, 16 (3) {[}b{]}, 24 (6) {[}c{]}\tabularnewline
\hline 
 & 5/2, 3/2 & 0.1411 (29) &  &  &  & \tabularnewline
\hline 
 & 5/2, 1/2 & 0.2276 (29) &  &  &  & \tabularnewline
\hline 
\end{tabular}
\par\end{centering}

\caption{{\label{tab:HFS}Measured level splittings and the hyperfine constants determined. Statistical uncertainty estimates are given for the measurements. Literature values reproduced from ref. \cite{D_Amico_1999}, table II, with {[}a{]} D'Amico et. al. (1999) \cite{D_Amico_1999}, {[}b{]} Jackson \& Coulombe (1972) \cite{Jackson_1972}, {[}c{]} Fischer et. al. (1974) \cite{Fischer_1974}.}} 
\end{table*}

\subsection{Quantum beat imaging\label{sect:QBI}}

\subsubsection{Phenomenology}

For a quantum beat imaging methodology, the full $(x,y,t)$ data volume can be transformed directly to the frequency domain via application of an FT to each pixel. This is a different approach from the treatment of the $\beta_{L}(t)$ parameters detailed above, since it does not require inversion of the raw image stack, hence is directly applicable even in the case of complex, non-invertable images (see Sect. \ref{sec:dataAnalysis}). The resulting $(x,y,\nu)$ images provide the Fourier components of the time-resolved photoelectron images.\footnote{By analogy with other imaging techniques, this can also be termed ``hyperspectral" VMI.} In cases where inversion is applicable, the processed data volume $(x_i,y_i,t)$ can be similarly transformed.

In the present case, the frequency components correspond to the hyperfine level splittings, and there is separation of the $I=0$ isotopes from the $I\neq 0$ isotopes in the frequency-domain, since only the latter can contribute to the time-dependence of the signal (i.e. $\nu >0$), while the former will only contribute to the time-independent (DC) part of the signal (i.e. $\nu =0$). Furthermore, if the Fourier components are uniquely associated with a particular isotope, then different frequency-domain images will correspond to isotopically-resolved wavepacket components and associated frequency-domain photoelectron images. For this assertion to be valid, the isotope signals must be incoherent, and the level-splittings must be resolved in the frequency-domain. If these conditions hold, then the Fourier components provide a means to obtain isotopically-resolved photoelectron distributions, correlated with pairs of hyperfine states. (This is conceptually similar to the recent photoelectron-photoion coincidence measurements mentioned previously \cite{O_Keeffe_2013,Gryzlova_2015}, except that the information obtained is a set of Fourier image components from a dynamical system, rather than state resolved photoelectron distributions.)



\begin{figure*} 
\begin{center}
\includegraphics{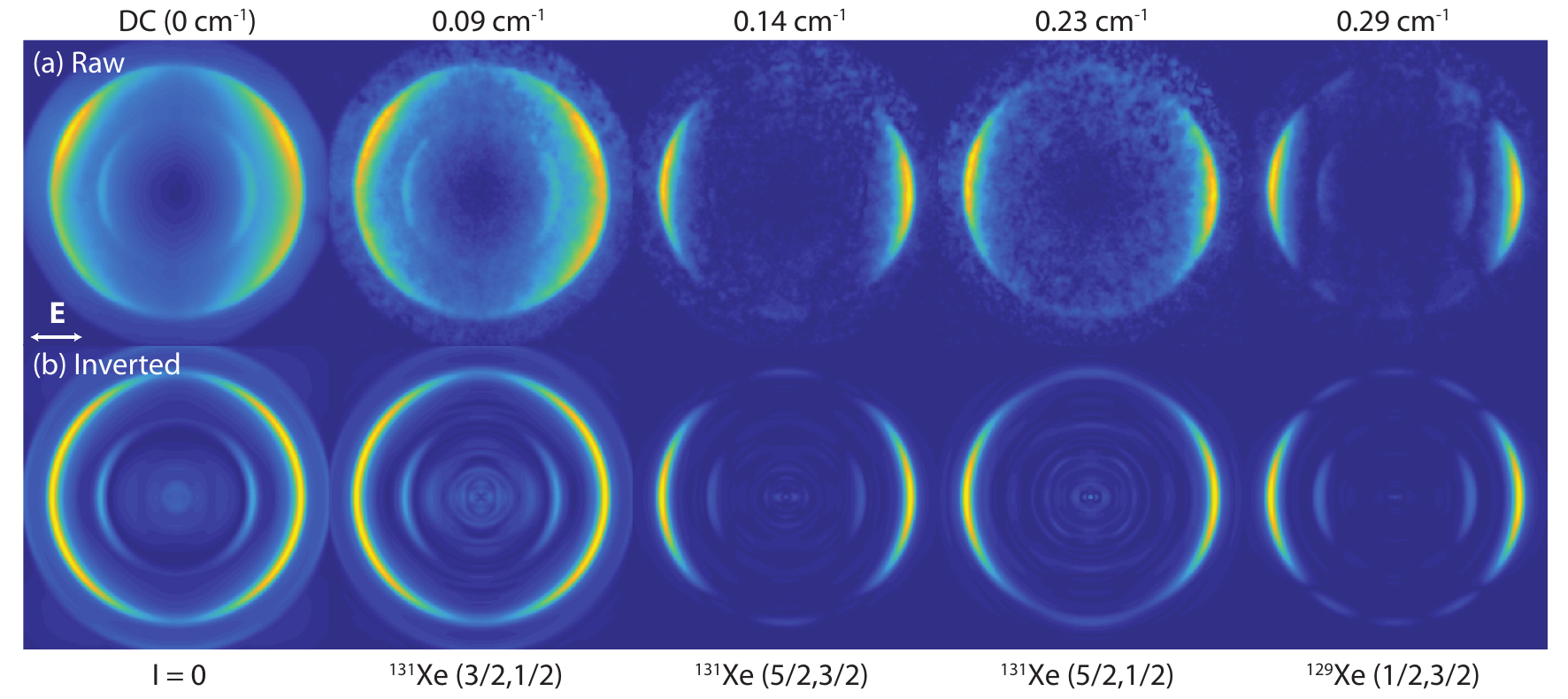}
\caption{{Fourier image components (absolute value) corresponding to the observed
features (see fig.~{\ref{586171}} and
table~{\ref{tab:HFS}}). (a) FT images from the raw
image stack. (b) FT images from the inverted image stack. Full image
stacks are shown in fig.~{\ref{701979}}.~ Colour maps
were normalised independently for each image.
{\label{906292}}%
}}
\end{center}
\end{figure*}

\begin{figure*} 
\begin{center}
\includegraphics{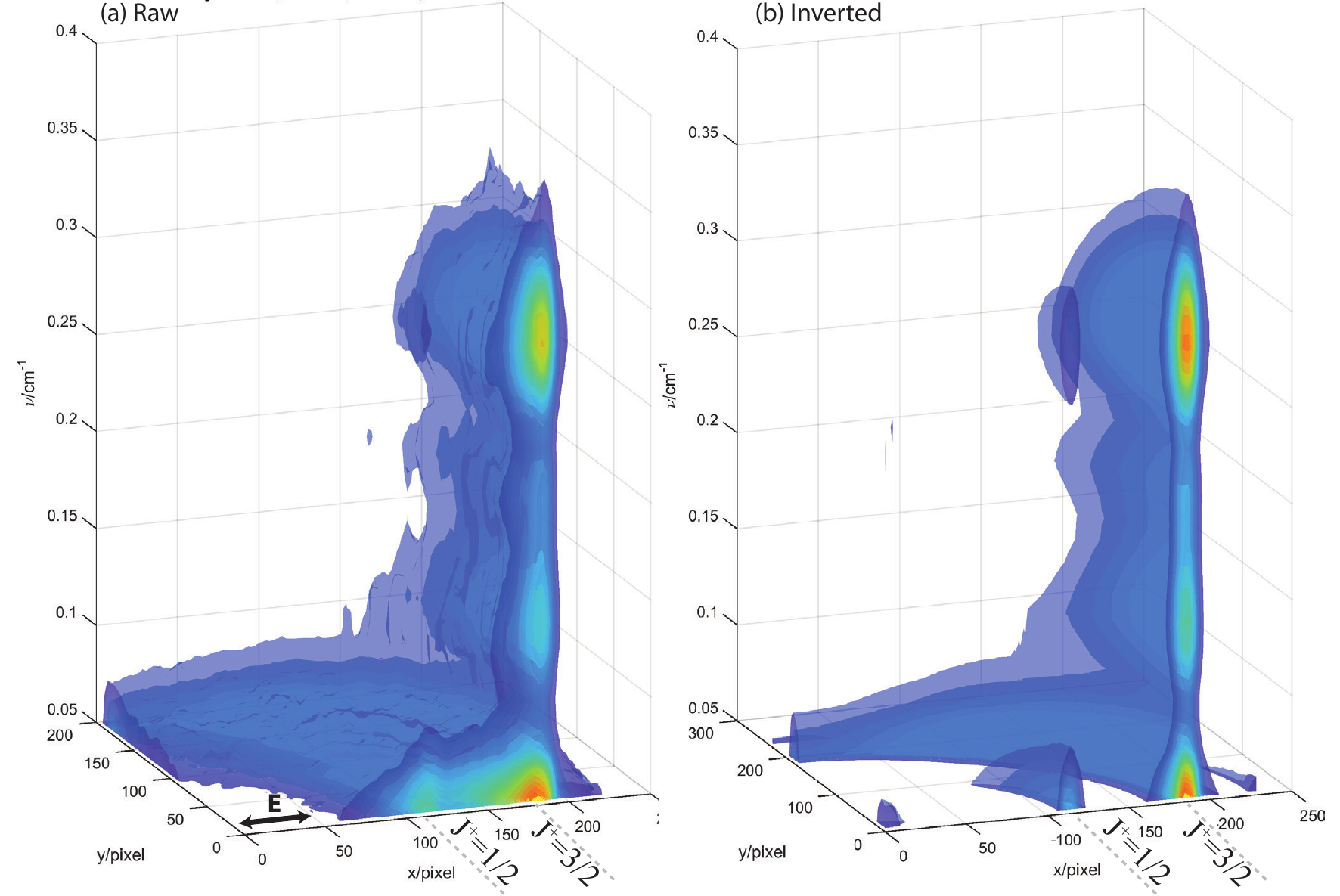}
\caption{{Fourier image volumes (absolute values, single quadrant). (a) FT of the
raw image stack~\(\left(x,y,\nu\right)\). (b) FT of the inverted image
stack~\(\left(x_i,y_i,\nu\right)\). Isosurfaces show 10 - 90\% photoelectron
signal. The full data volume was down-sampled and smoothed for these
renderings. The main features correspond to ~the images shown in
fig.~{\ref{906292}}. ~~
{\label{701979}}%
}}
\end{center}
\end{figure*}

Figures \ref{906292} and \ref{701979} illustrate the Fourier domain images (absolute values). The main discrete frequency components, correlated with the level splittings given in table \ref{tab:HFS}, are shown in Figure \ref{906292}; Figure \ref{701979} provides renderings of the full frequency domain image volumes. A number of phenomenological observations may be made from these results. 

In both sets of images, the angular features are peaked along the laser polarization axis ($x$-axis, corresponding to positive $\beta_2$), and this appears to be the dominant contribution in all cases. In terms of the hyperfine wavepacket, the images correlate with beat frequencies, hence pairs of $F$ levels. The photoelectron interference pattern for each pair therefore contains two contributions \cite{Chien1983}: 
\begin{enumerate}
\item the ``intrinsic" photoionization interferences, due to the partial-wave composition of the continuum wavefunction, which would be observed for ionization of a single (eigen)state; 
\item additional interferences which arise in the sum over pairs of $F$ states, and include a time-dependence due to the time-evolution of the prepared wavepacket. 
\end{enumerate}



In the case of $^{131}$Xe, there is an additional layer of complexity, since there are multiple $F$ pairs which contribute to the final photoelectron images. In these results, it appears that interferences between the components\footnote{More precisely, between the underlying photoelectron wavefunctions correlated with each of the three $F$ states involved.} at 0.09 and 0.14~cm$^{-1}$ - which have lobes with different angular spreads (see fig. \ref{906292}) - is the main source of the $\beta_4(t)$ oscillations observed 
(Figs. \ref{622685} \& \ref{586171}); this is also consistent with the lack of a $\beta_4(\nu)$ feature at 0.09~cm$^{-1}$ (Fig. \ref{586171}). In contrast, for $^{129}$Xe only a single Fourier image component is present (0.29~cm$^{-1}$), indicating that any $\beta_4(t)$ oscillations correlated with this isotope originate from just this $F$ pair; this is consistent with the significant $\beta_4(\nu)$ feature observed at this frequency (fig. \ref{586171}). This is also suggested by the greater angular complexity in the images (higher $L$ terms), as compared to the $^{131}$Xe images. This is observed most clearly in the inverted image, which contains multiple lobes for both spectral features. Of note here is the fact that, in the Fourier domain images, higher order angular patterns than usually allowed for a 2-photon process ($L_{max}=4$) can appear due to the differential nature of the images. The images in this case therefore indicate that the intrinsic ionization dynamics of the two $F$ states involved would result in different four-fold distributions. However, quantitative analysis along these lines requires knowledge of the photoionization matrix elements, either from calculation or via retrieval from experimental data. In the latter case, previous work has shown that this may be possible for hyperfine wavepacket data \cite{Strand1978,Hansen1980,Chien1983}, and the data obtained in this work may also contain sufficient information for such a retrieval. Work is ongoing in this direction, and some general comments are presented in the following section.

Another interesting feature of the images presented in fig. \ref{906292} is the absence of any appreciable intensity in the inner ring for the images correlated with 0.14~cm$^{-1}$ and 0.23~cm$^{-1}$. This indicates very little population of the $J^+=1/2$ state of the cation via ionization from these components of the hyperfine wavepacket. 
Finally, it is interesting to note that the widths (radial spread) of the features is not constant. This is seen most clearly in the inverted image volumes of fig. \ref{701979}, but is also apparent in the inverted images of fig. \ref{906292}. In particular, the 0.29~cm$^{-1}$ feature has a larger radial extent that the 0.14~cm$^{-1}$ feature, indicating slight changes in the photoelectron energy spectra associated with these components. 


\subsubsection{Phase imaging and wavepacket treatment}

\begin{figure*} 
\begin{center}
\includegraphics{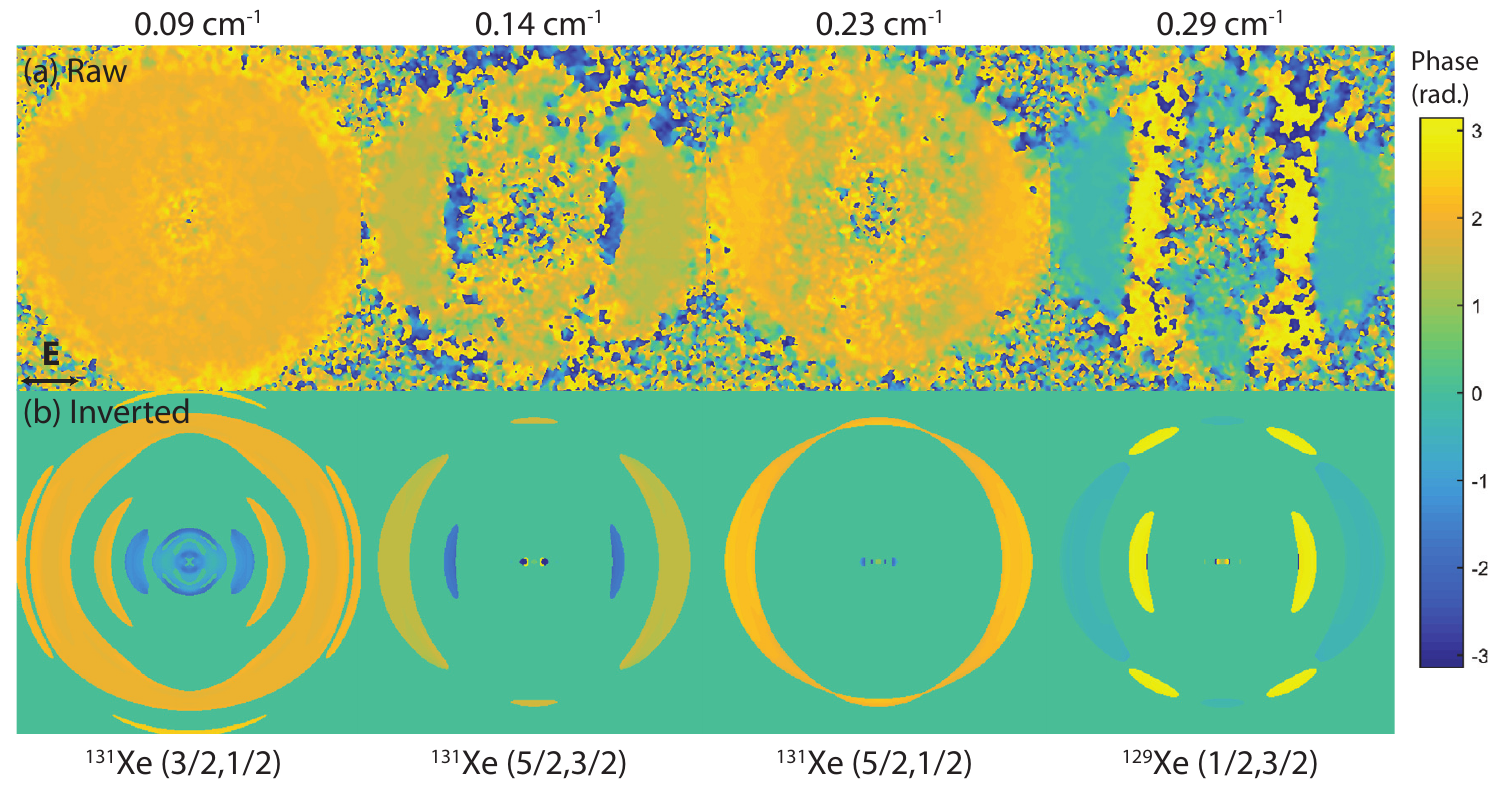}
\caption{{Fourier image components (phases) corresponding to the observed features
(see fig.~{\ref{586171}} and
table~{\ref{tab:HFS}}). (a) FT images from the raw
image stack. (b) FT images from the inverted image stack. The
corresponding absolute value images are presented in
fig.~{\ref{906292}}.
{\label{765630}}%
}}
\end{center}
\end{figure*}

In the preceeding discussion, the QB imaging results were presented along with a basic phenomenological discussion. Further insight into the results can be obtained via the associated phase structure of the $(x,y,\nu)$ data. Fig. \ref{765630} shows the phase images for the observed features, for both the raw and inverted images, corresponding to the absolute value images shown in fig. \ref{906292}. Broadly speaking, the phase structure is similar for both the raw and inverted images, indicating the possibility of accessing phase information directly from the raw image stacks as suggested previously. Unsurprisingly, the inverted images contain a clearer and more pronounced phase structure\footnote{It is of note here that the inverted phase images include an intensity mask, with a threshold at $\approx$5\% of the photoelectron yield, to remove spurious phase noise in energy regions with no signal.}. This occurs simply because the photoelectron features do not overlap in the inverted (slice) images, which show clear rings and associated phases, in contrast to the raw (crush) images, which show continuous phase structure over the images.

Exploration of the phase structure in the images provides an alternative analysis methodology, which can be compared with many of the observations previously discussed. For instance, the inner and outer photoelectron bands are out-of-phase; this effect is particularly clear in the 0.14 and 0.29~cm$^{-1}$ features, and is in agreement with the phase shift observed in the $\beta_L(t)$ plots pertaining to the two different final states (fig. \ref{622685}). The appearance of additional angular structure in the 0.29~cm$^{-1}$ feature, discussed in the previous section as a result of different final continuum states related to each ionizing $F$ state, is now made clearer by the alternating phase pattern in the outer ring. This phase structure indicates the phase difference between the two $F$ state contributions, with the equatorial region shifted from the four-fold lobe structure by just over $\pi$ radians. This phase structure quantitatively maps the change in angular structure as a function of $t$ observed in the original time-domain images (fig. \ref{772944}).

A formal statement of the phase contributions can be obtained from a wavepacket treatment, in which the full complexity of the photoionization dynamics remains implicit. In general, the signal for a PAD measured by one-photon ionization with linearly polarized light from a wavepacket can be written as 
\begin{equation}
I(\theta,E,t)\propto| \langle \Psi(t) |\mu_Z|\Phi_f^+\psi_e(E,\theta)\rangle |^2.
\label{eq:tpad}
\end{equation}

Here $\Psi(t)$ is the bound wavepacket, $\mu_Z$ is the dipole moment along the polarization direction of the light, $\Phi_f^+$ and $\psi_e(E,\theta)$ are the final ionic state and photoelectron wavefunctions respectively. To begin with assume that the wavepacket is a superposition of two hyperfine components,

\begin{equation}
\Psi(t)=c_{1}e^{iE_{1}/\hbar t}\left|F_{1}\right>+c_{2}e^{iE_{2}/\hbar t}\left|F_{2}\right>.
\end{equation}

where $c_{i}$ and $E_{i}$ are the coefficients and energies of the hyperfine states. Inserting this into into Eq.~\ref{eq:tpad} and expanding gives,

\begin{eqnarray}
I(\theta,E,t) & = & P_{1}|d_{1}^{f}|^2+P_{2}|d_{2}^{f}|^2 \nonumber  \\
 & + & c^*_{1}c_{2}d_{1}^{f}d_{2}^{f*}e^{i\Omega t}+c_{1}c^*_{2}d_{1}^{f*}d_{2}^{f}e^{-i\Omega t}.
\end{eqnarray}

$P_{1}=|c_{1}|^2$ and $P_{2}=|c_{2}|^2$ are the populations of the two hyperfine states. $d_{1}^f=\langle F_{1}|\mu_Z|\Phi_f^+\psi_e(E,\theta)\rangle$ is the ionization dipole matrix out of the $F_{1}$ state, and $d_{2}^f$ is the same out of $F_{2}$. The dependence of these on $E$ and $\theta$ is implicit. $\Omega=(E_{2}-E{1})/\hbar$ is the angular beat frequency between the hyperfine states. Note that the last two terms in the above expression are conjugates. As a result, if the dipole matrix elements and wavepacket coefficients are expressed in terms of their amplitudes and phases, the following expression results -

\begin{eqnarray}
I(\theta,E,t) & = & P_{1}|d_{1}^{f}|^2+P_{2}|d_{2}^{f}|^2 \nonumber \\
 & + & 2|c_{1}||c_{2}||d_{1}^{f}||d_{2}^{f}|\cos\left[\Omega t+\Delta\phi^c_{1,2}+\Delta\phi_{1,2}^f\right].
\label{eq:2state}
\end{eqnarray}

From this expression the images obtained by the FT of $I(\theta,E,t)$ can be understood. 

The $I(\theta,E,\nu=0)$ image corresponding to the DC component of the FT is given by $P_{1}|d_{1}^{f}|^2+P_{2}|d_{2}^{f}|^2$. The angle and energy dependent structure in these images is therefore entirely determined by the magnitudes of the ionization dipoles out of each component states. The contribution of each component state is determined by their populations. These images will have only a real component, and reflect the ``intrinsic" photoionization dynamics, hence the sum over the corresponding PADs from each component state.

The $I(\theta,E,\nu)$ images corresponding to the beat frequencies on the other hand should have an amplitude and phase. The structure in the amplitude image is determined by the product of the ionization dipoles - $|d_{1}^{f}(E,\theta)||d_{2}^{f}(E,\theta)|$ - and that of the phase image by the phase difference between the ionization dipoles - $\Delta\phi_{1,2}^f(E,\theta)$. Note both the amplitude and phase are offset by constants determined by the product of the magnitudes of the wavepacket coefficients, and their phase differences ($\Delta\phi^c_{1,2}$) respectively. As such these images represent a direct measurement of the interfering ionization pathways out of the component states of the wavepacket. Eq~\ref{eq:2state} can be easily generalized to a wavepacket of $N$ component eigenstates -

\begin{eqnarray}
I(\theta,E,t) & = & \sum_{i}P_{i}|d_{i}^{f}|^2+\sum_{j< k}2|c_{j}||c_{k}||d_{j}^{f}||d_{k}^{f}| \nonumber \\
 & \times & \cos\left[\Omega t+\Delta\phi^c_{j,k}+\Delta\phi_{j,k}^f\right].
\label{eq:Nstate}
\end{eqnarray}

The indicies $i$,$j$ and $k$ run over all $N$ states. Thus in the case where numerous eigenstates compose the wavepacket, the FT images for each beat frequency separate out the interference patterns resulting from ionization from each pair of states.

\section{Conclusions and future work}

In this work, the combination of a 130~nm VUV source with a pump-probe methodology and photoelectron imaging measurements has been demonstrated, and used to probe a hyperfine wavepacket in Xe. The measurement of images over a large time-window, and with high timing accuracy, provided hyperfine splittings and coupling constants which were compared with literature values. The full imaging data was also investigated in the frequency domain, and the retrieval of images correlated with different isotopes and 
wavepacket components, via FT of the image volume, was demonstrated and explored.

From a QBS perspective, the use of a fs VUV source (ideally tuneable), combined with a UV ionization probe, provides a method applicable to a range of systems, and which can be used to interrogate high-lying manifolds which are typically hard to access \cite{Paul2005}. As discussed in ref. \cite{Sukhorukov2012}, ``in view of the paucity of the data measured so far and because of the possibility of carrying out accurate calculations, PADs from excited states of the heavier rare-gas atoms offer an interesting opportunity for combined experimental and theoretical studies in the future"; the authors also mention that ``PAD studies involving selected excited states of the rare-gas atoms are also scarce", as are studies of auto-ionizing states.  Ultrafast pump-probe VUV-UV measurements certainly provide a technique suitable for investigating this area, and the broadband laser pulses are also suitable for cases with larger level splittings, hence QBS of lighter elements. For an extended discussion of recent developments in precision laser spectroscopies in the VUV and XUV, see ref. \cite{Eikema2011}.
For short pulses, limitations on frequency resolution are placed, effectively, only by the temporal sampling parameters of the measurements. The difficulty of such measurements lies, instead, in the lengthy experimental runs that may be required for very high-precision measurements, and consequent requirements for long-term experimental stability. The ability to extract frequency-correlated images is a novel emergent property of time-resolved photoelectron imaging experiments with sufficient spatio-temporal sampling (data volume), and one which allows for beat-component and isotope-correlated imaging in favourable cases, as demonstrated herein.

From a more general time-domain spectroscopy perspective, the results presented herein are a subset of wavepacket measurements. In this vein, there is a significant literature on time-resolved photoelectron spectroscopy and imaging, PADs and related work (see, for instance, ref. \cite{Wu_2011} for a basic introduction, and refs.  \cite{Suzuki_2001,Seideman_2001,Suzuki_2006,Stolow_2008,Reid_2012,Hockett2018,Hockett2018a} for further discussion and review). Despite this extant work, a Fourier transform photoelectron imaging methodology has not previously been explored, to the best of our knowledge. Since the analysis methodology is somewhat obvious, in the sense that it is a clear extension of standard (but lower dimensionality) time-domain analyses, this is presumably due to the lack of sufficient data volumes in other cases. Many factors may contribute here, including the native time-scale and complexity of the process under study, the experimental difficulty of obtaining sufficient measurements, or additional sampling constraints. Despite these challenges, the ability to obtain some phase information directly from the raw images is, potentially, an interesting feature of this type of analysis and may motivate future studies.

Apart from the image processing, and determination of the hyperfine coupling constants, the analysis presented herein is relatively phenomenological. As mentioned above, future work will aim to address this point by consideration of both the hyperfine wavepacket and photoionization dynamics in more detail, making use of the formalism developed by Berry and coworkers \cite{Chien1983}. Such investigation is, necessarily, rather involved, but would provide a more detailed insight into the import of the FT images presented herein, and the information content of the quantum beat imaging methodology. In the case of a sufficiently high information content, it is possible that the full photoionization dynamics and wavepacket dynamics could be retrieved/reconstructed from the experimental data - this is a form of ``complete" experiment in photoionization terminology, and can also be considered as a form of quantum metrology or tomography in the language of quantum information \cite{Kleinpoppen_2013,Hockett2018,Hockett2018a}.


\section{Acknowledgements}
We are grateful to Andrey Boguslavskiy, Denis Guay and Doug Moffat for assistance with the experimental infrastructure, and technical support. AS thanks the NSERC Discovery Grant program for financial support.

\section{Supplementary Material\label{sec:SuppMat}}

Raw data, processing routines and additional analysis documentation are available via an online OSF repository, \href{https://osf.io/ds8mk/}{\textbf{Quantum Beat Photoelectron Imaging Spectroscopy of Xe in the VUV}}, DOI: \href{http://dx.doi.org/10.17605/OSF.IO/DS8MK}{10.17605/OSF.IO/DS8MK}. This repository provides a complete reporting of the analysis routines (Matlab scripts), including full details of the image processing and Fourier transform routines (which included zero-padding and a Hann window function in the results shown herein), for readers who wish to explore the technique in further detail or build on the code-base developed.

\clearpage
\bibliographystyle{apsrev4-1}
\bibliography{converted_to_latex}

\end{document}